\documentclass[review]{elsarticle}

\usepackage{lineno,hyperref}
\usepackage{siunitx}
\usepackage{amsmath}
\usepackage{latexsym}
\usepackage{amssymb}

\modulolinenumbers[5]

\journal{Infrared Physics and Technology}

\begin{document}

\begin{frontmatter}

\title{Ultra High Molecular Weight Polyethylene: optical features at millimeter wavelengths}

\author[a]{G. D'Alessandro} 
%\ead{giuseppe.dalessandro@roma1.infn.it}
\author[a]{A. Paiella}
\author[a]{A. Coppolecchia}
\author[b]{M.G. Castellano} 
\author[c]{I. Colantoni} 
\author[a]{P. de Bernardis} 
\author[a]{L. Lamagna} 
\author[a]{S. Masi}

%\address[a]{Physic Department, University Sapienza of Rome, 00185 Piazz. Aldo Moro 5 Roma, Italy}
%\address[b]{IFN - Institute for photonics and nanotechnologies, 00156 Via Cineto Romano 42,  Roma, Italy} 

\address[a]{Dipartimento di Fisica, Universit\`a di Roma La Sapienza , P.le A. Moro 2, 00185 Roma, Italy}
\address[b]{Istituto di Fotonica e Nanotecnologie - CNR, Via Cineto Romano 42, 00156 Roma, Italy}
\address[c]{Istituto di Fotonica e Nanotecnologie - CNR, Via Cineto Romano 42, 00156 Roma, Italy; Now at Dublin Institute for Advanced Studies, School of Cosmic Physics Astronomy and Astrophysics Section, 31 Fitzwilliam Place, D02 XF86, Dublin, Ireland}

\begin{abstract}
The next generation of experiments for the measurement of the Cosmic Microwave Background (CMB) requires more and more the use of advanced materials, with specific physical and structural properties. An example is the material used for receiver's cryostat windows and internal lenses. The large throughput of current CMB experiments requires a large diameter (of the order of 0.5m) of these parts, resulting in heavy structural and optical requirements on the material to be used. Ultra High Molecular Weight (UHMW) polyethylene (PE) features high resistance to traction and good transmissivity in the frequency range of interest. In this paper, we discuss the possibility of using UHMW PE for windows and lenses in experiments working at millimeter wavelengths, by measuring its optical properties: emissivity, transmission and refraction index. Our measurements show that the material is well suited to this purpose.\\
\end{abstract}


\end{frontmatter}

\section{Introduction}

The survey sensitivity of CMB experiments is currently pursued increasing the number of radiation modes detected, either using large-format arrays of single-mode detectors or using arrays of multi-moded detectors. In both cases, the large optical throughput of the instrument requires large optical elements. The first element (skywise) of the receiver optical train is the cryostat window. Its optimization is crucial because it represents the first filter for the radiation, therefore must have a high transmission in the frequency bands of the experiment and, at the same time, must withstand without bending too much the large inwards force due to external pressure and internal vacuum.  

Nowadays, High Density Polyethylene (HDPE) is widely used for cryostat windows \cite{OLIMPO2003,Coppoproc,debe12a,prism,blast,spider} and lenses in the millimeter-waves frequency range of interest here \cite{spider,2014A&A...565A.125S, swipe,ACT_spie}. UHMW PE is a kind of thermoplastic polyethylene. It has extremely long chains of polyethylene, all aligned in the same direction. The molecule forming UHMW chains is heavier than that of HDPE, and this makes the UHMW stronger and easier to be machined \cite{SUI2009404,ALCOCK2006716}. The tensile strength of UHMW PE is twice that of HDPE, therefore it represents a good candidate for replacing HDPE. 

Although some experiments devoted to the measurements of the CMB, like the Atacama Cosmology Telescope, CLASS, and BRAIN pathfinder, already used this material for their optical components \cite{ACT_spie,MUSIC_spie, LMT, brain, class}, direct measurements of its most important optical characteristics cannot be found in the literature yet. In this paper we report measurements of emissivity, transmission and refraction index for UHMW PE, providing useful data for the design of new CMB experiments and other optical instruments for mm-wave measurements. 

\section{Mechanical characteristics and simulations}\label{sec:mec}

\begin{figure}[h]
\begin{center}
{\includegraphics[scale=0.25]{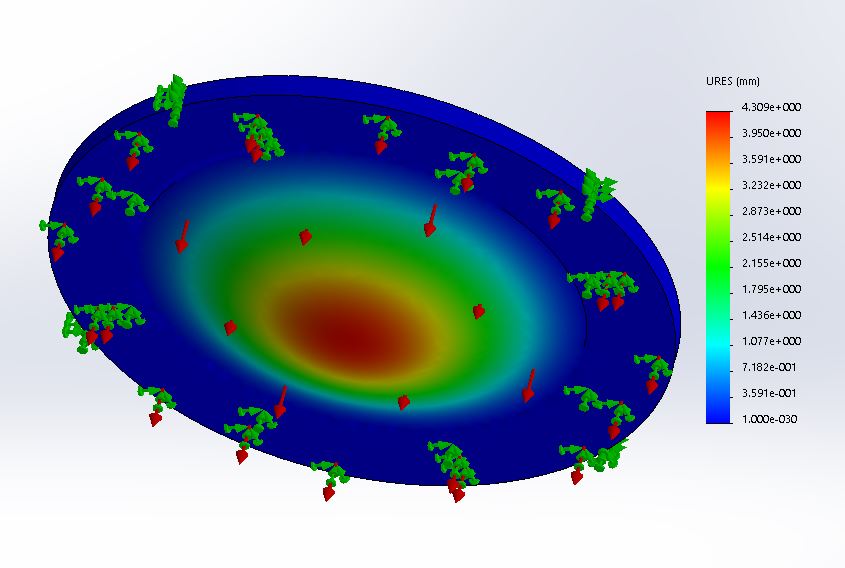}}
{\includegraphics[scale=0.25]{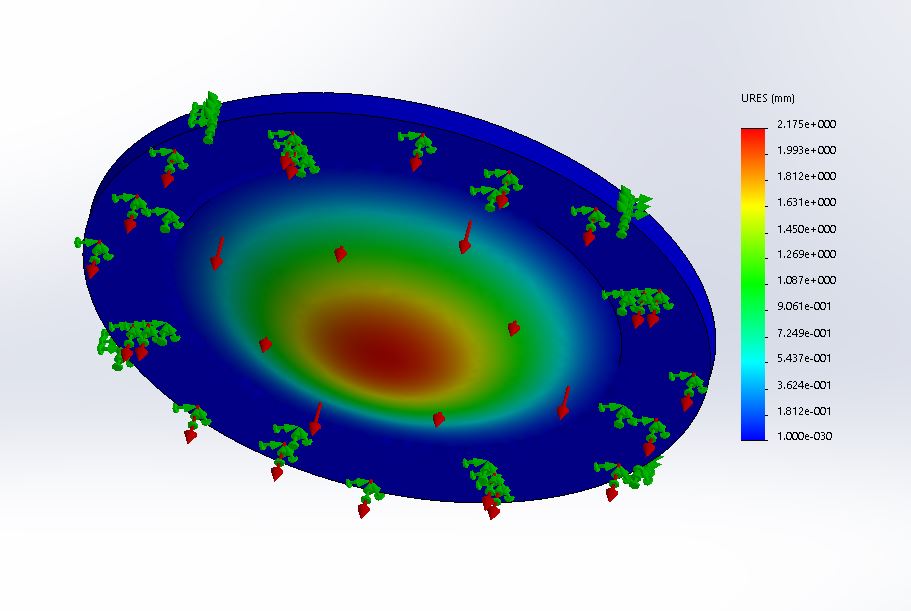}}
\caption{Finite elements simulation performed with ANSYS. The simulated window has a  diameter of \SI{50}{cm}, a thickness of \SI{2.5}{cm}, and its temperature has been set at \SI{300}{K} (room temperature). The windows are subjected to pressure of \SI{1000}{mbar} and show a displacement of \SI{4.3}{mm} for the HDPE and \SI{2.2}{mm} for the UHMW}
\phantomsection\label{simulaz}
\end{center}
\end{figure}

The tensile strength of UHMW PE is $\sim$ \SI{40}{MPa}, while that of HDPE is $\sim$ \SI{20}{MPa}. In order to evaluate the impact of this characteristics on the mechanical performance of large windows, we performed a load simulation of a large diameter window, for both materials, using a finite element simulator\footnote{Solidworks in this case, but other software, such as ANSYS and COMSOL MULTIPHYSICS give the same results}. The simulated window has a  diameter of \SI{50}{cm}, a thickness of \SI{2.5}{cm}, and its temperature has been set at \SI{300}{K} (room temperature). The pressure difference from top to bottom is \SI{1000}{mbar}\footnote{Usually the internal cryostat pressure is $\sim$ $10^{-6}mbar$}.
As expected, the simulations \figurename~\ref{simulaz} confirm that the UHMW window deformates less. This means that it is possible to use less material in order to have the same mechanical performance of the HDPE, see Tab.~\ref{tab:defor}. Less material means less weight, and better transmission. Thanks to this mechanical feature, each secondary machining on the surface, such as anti-reflection coating, is easier and the risk of damage to the material is reduced.   

\subsection{Deformation vs. thickness simulations}
We use ANSYS software to show the deformation for three different cryostat windows, with diameter of \SI{10}{cm}, \SI{25}{cm} and \SI{50}{cm}, as a function of thickness. We obtain results showed in Tab.~\ref{tab:defor}.   

\begin{table}[h]
\centering
\begin{tabular}{c|c|c|c}
\hline
\hline
diameter  & thickness  & deformation UHMW& deformation HDPE  \\
$\left[\SI{}{cm}\right]$ & $\left[\SI{}{cm}\right]$ & $\left[\SI{}{mm}\right]$ &$\left[\SI{}{mm}\right]$  \\
\hline
\hline
50 & 1.0 & 14 & 19\\
50 & 2.0 & 4.0 & 7.9\\
50 & 2.5 & 2.2 & 4.3\\
\hline
25 & 1.0 & 2.0 & 3.9\\
25 & 2.0 & 0.3 & 0.6\\
25 & 2.5 & 0.2 & 0.3\\
\hline
10 & 1.0 & 0.07 & 0.1\\
10 & 2.0 & 0.01 & 0.03 \\
10 & 2.5 & 0.003 & 0.02\\
\hline
\hline
\end{tabular}
\caption{Displacements evaluate by using finite element simulations for three different cryostat window, subjected to \SI{1000}{mbar} of pressure as a function of thickness.}
\phantomsection\label{tab:defor}
\end{table}

The results in table shows how it is possible to reduce the thickness of windows from \SI{25}{mm} to \SI{20}{mm}, for each diameter, obtaining the same displacement and increasing the transmissivity, as it showed in \figurename~\ref{f1}.    

\section{Emissivity}
The emissivity of the optical components is crucial because it represents part of the background power incident on the detectors and part of the optical load on the fridge of the cryogenic system.

The design of the detectors is strictly correlated with the background power, as well as the performance of the fridge depends on the optical load. A typical emissivity for the optical elements working at millimeter wavelengths is around few percent \cite{2013InPhT..58...64S, bock1995emissivity, Halpern86}. The direct measurement of the emissivity is not simple, since the quantity which we want to measure is small and it is easy that the experimental setup is dominated by systematics. 

\begin{figure}[h]
\begin{center}
{\includegraphics[scale=0.04]{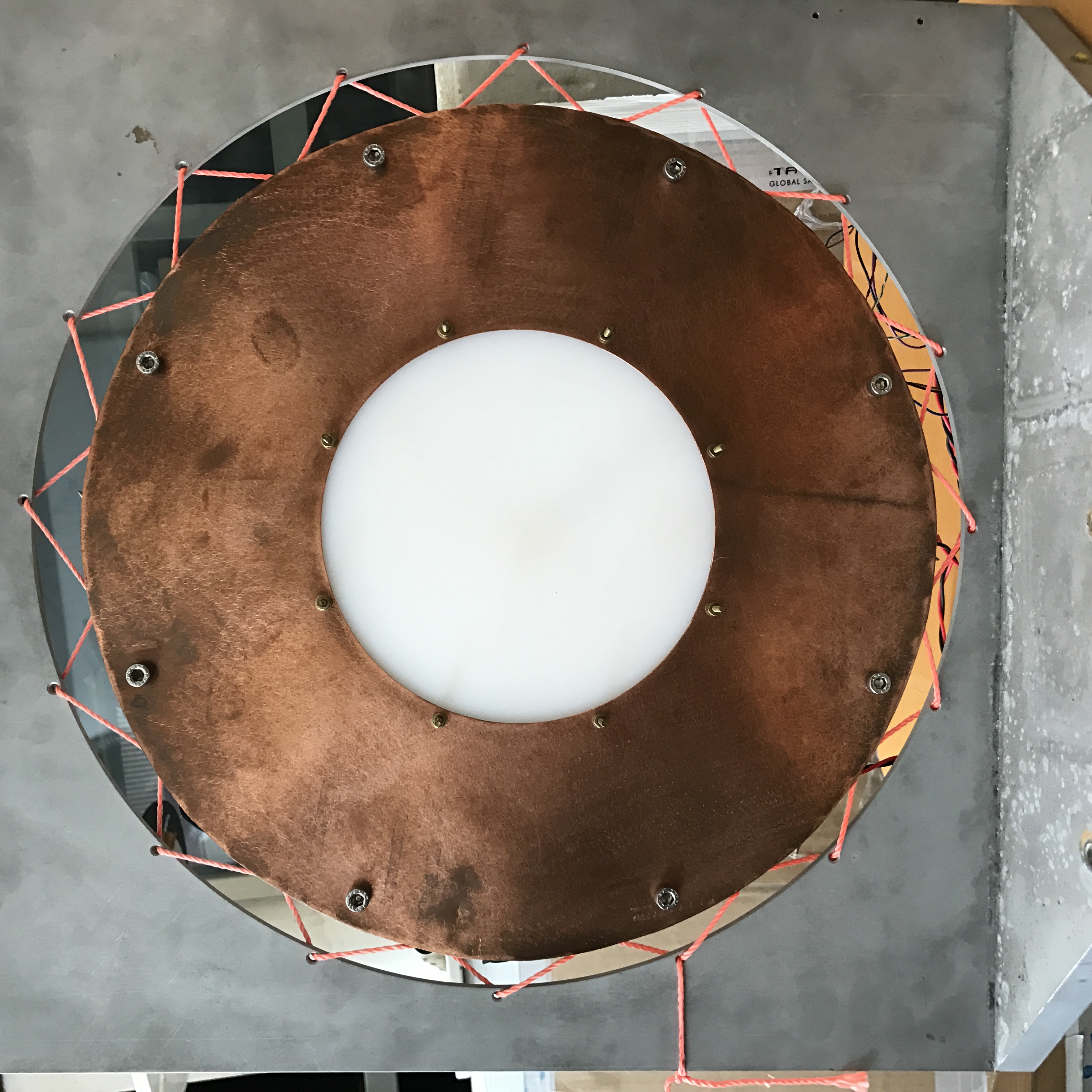}}
{\includegraphics[scale=0.04]{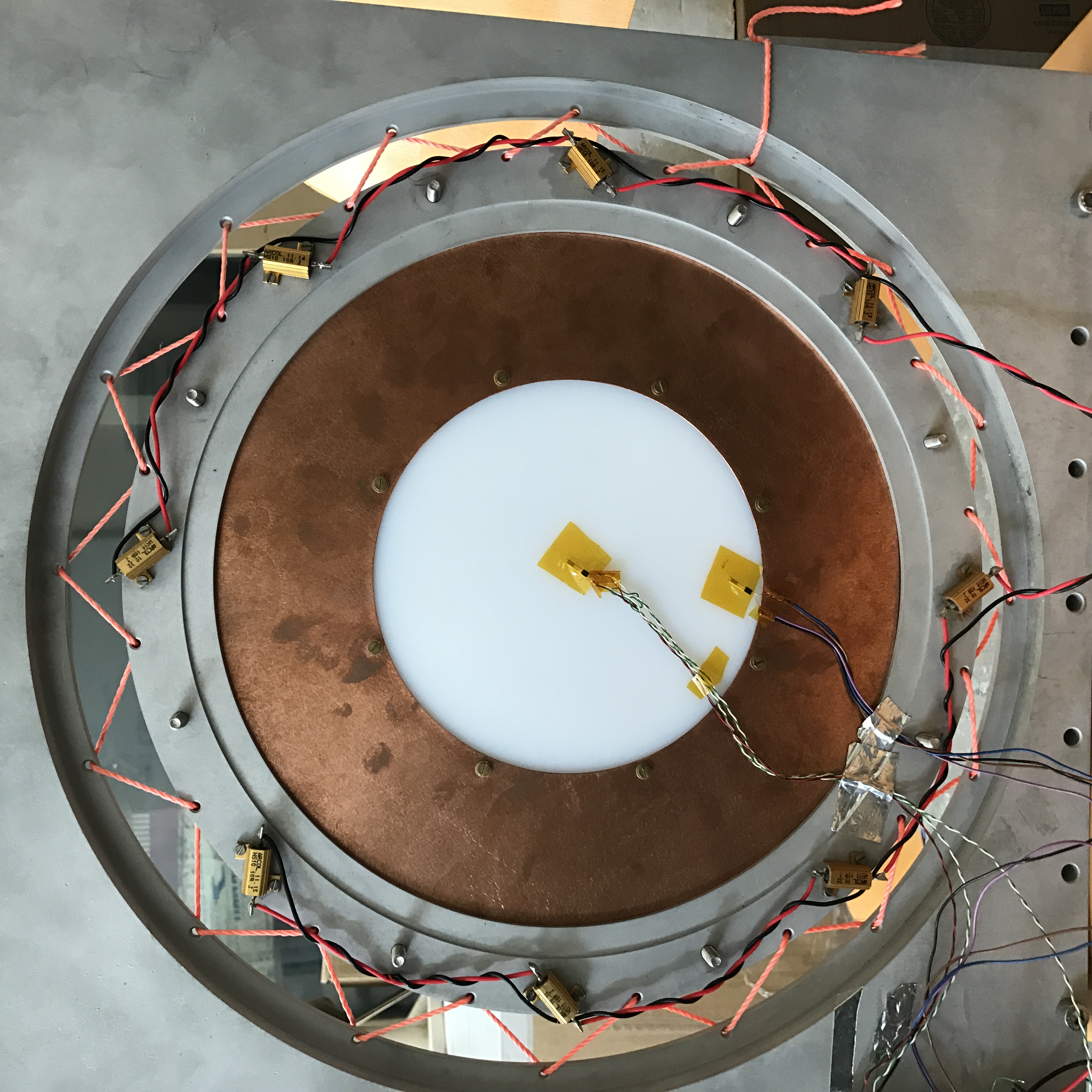}}
\caption{The UHMW disk is surrounded by a copper crown, needed the improve the thermalization. The copper crown is surrounded by an aluminum ring where a number of heathers is placed. The heaters allow to heat the UHMW and the temperature is monitored by two thermometers: one located at the center and the second located on the edge of the UHMW disk. The aluminum ring, to which the copper crown and the UHMW disk are anchored, is suspended in air through some kevlar fibers.}
\phantomsection\label{photo}
\end{center}
\end{figure}

In order to estimate the UHMW emissivity, we build a setup like in \cite{2013InPhT..58...64S}: a disk of UHMW, thickness of \SI{10}{mm}, surrounded by a copper crown, which is, in turn, surrounded by an Aluminum ring, is suspended in air through some kevlar fibers, in order to thermally insulate the UHMW from its metal support. The aluminum ring is equipped with a number of evenly spaced heaters, which allow to control the temperature, while the disk of UHMW is equipped with two thermometers PT100: one at the center and one near the copper crown. \figurename~\ref{photo} shows this setup. We verified, with dedicated tests, that the PT100 which is placed at the center of the UHMW disk, does not corrupt the measurement.

\subsection{Instrumental setup}
For this measurement, the instrumental setup is composed of low temperature detectors, in particular kinetic inductance detectors (KIDs) \cite{Mazin, Doyle}, and therefore of a cryogenic system, equipped with a dedicated optical system, and the readout electronics.

\begin{figure}[h]
\begin{center}
\includegraphics[scale=0.5]{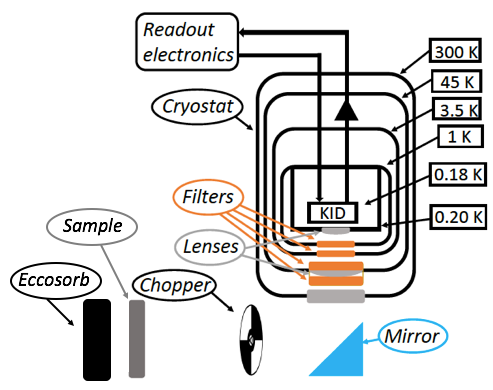}
\caption{Scheme of the measurement setup used for the emissivity. The signal coming from the sample is chopped with a blackbody at \SI{300}{K} and reflected in the cryostat by a $45^{\circ}$ mirror. Inside the cryostat, the signal passes through the cryostat optical system and illuminates the detectors. The bias and the readout signals of the detector are monitored through a dedicated electronics.} 
\phantomsection \label{fig:schema_misure}
\end{center}
\end{figure}

The sample is placed in such a way that the signal coming from it is chopped with a blackbody at $300K$ (Eccosorb sheet) before entering in the cryostat, the Eccosorb sheet is glued to Aluminum sheet and the temperature is controlled with a PID system by using a PT100 thermometer and resistor like heather. The bias and the readout signals of the detector are monitored through a dedicated electronics, which include a frequency synthesizer, a signal splitter, an IQ mixer demodulator, a low noise amplifier, a warm amplifier, and an ADC. \figurename~\ref{fig:schema_misure} shows the scheme of this measurement setup. 

\subsubsection{Detectors}
The detectors consist of two single pixel KIDs, built on $\SI{1}{cm}\times\SI{1}{cm}$, $\SI{300}{\micro m}$ thick, high-quality (FZ method), intrinsic Silicon wafer, with high resistivity ($\rho>\SI{10}{k\ohm.cm}$) and double side polished. The \SI{90}{GHz} KID is a TiAl bilayer \SI{10}{nm} thick Titanium $+$ \SI{25}{nm} thick Aluminum, while the \SI{150}{GHz} KID is in Aluminum \SI{25}{nm} thick. For both the detectors, the feedline is a coplanar waveguide, matched to \SI{50}{\ohm}.

\begin{figure}[h]
\begin{center}
{\includegraphics[scale=0.22]{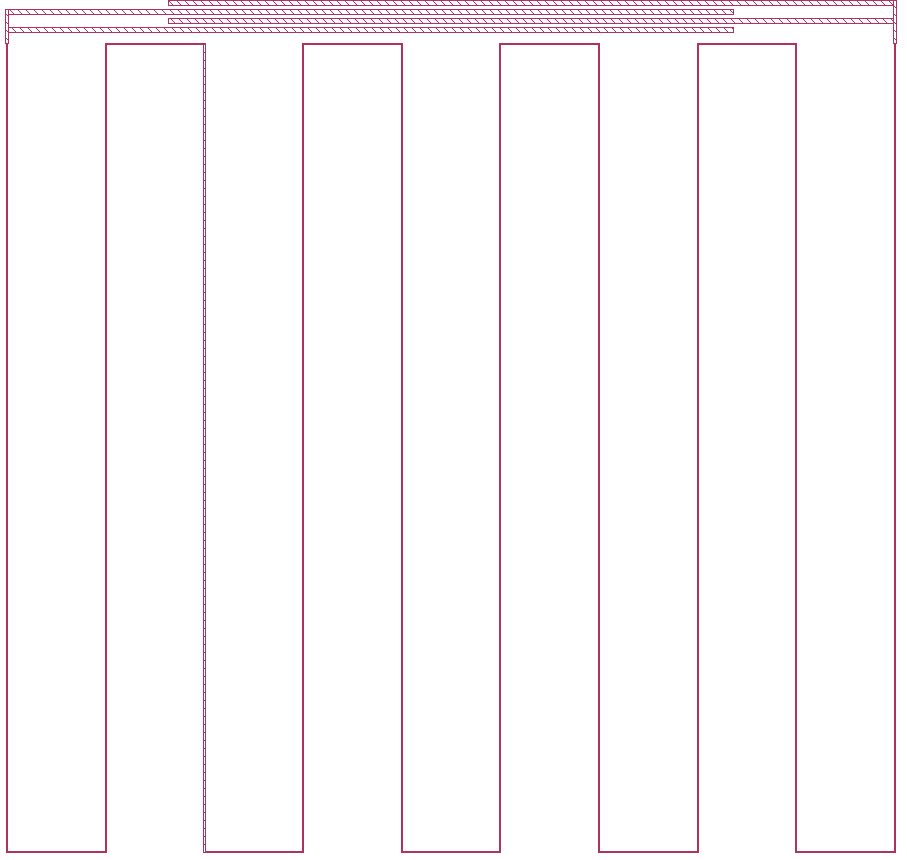}}\quad
{\includegraphics[scale=0.22]{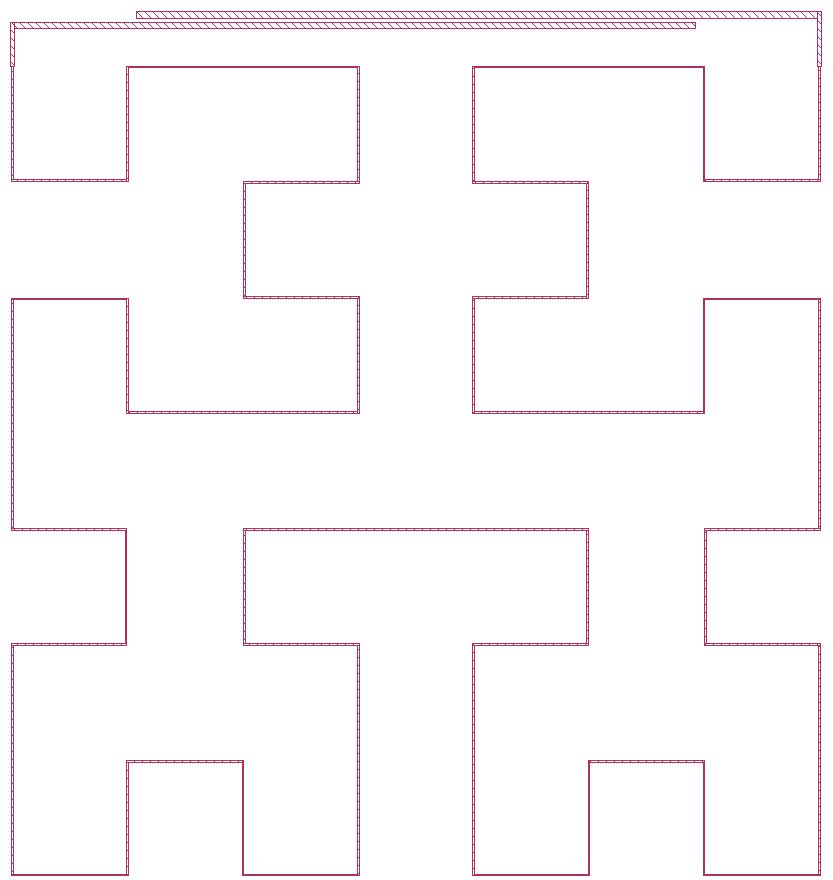}}
\caption{Designs of the \SI{90}{GHz} (\emph{left panel}) and of the \SI{150}{GHz} (\emph{right panel}) KID.}
\phantomsection \label{fig:design_KID}
\end{center}
\end{figure}

The absorber of the \SI{90}{GHz} KID is a standard meandered line, while that of the \SI{150}{GHz} KID is a III order Hilbert curve. For both the detectors, the capacitor has the interdigitated geometry, designed in order to guarantee the lumped condition, and to have a resonance frequency around \SI{1}{GHz} for the \SI{90}{GHz} KID, and \SI{2}{GHz} for the \SI{150}{GHz} KID. \figurename~\ref{fig:design_KID} shows the designs of the \SI{90}{GHz} and \SI{150}{GHz} KID.

The detectors are fabricated at the Istituto di Fotonica e Nanotecnologie (IFN) of the Consiglio Nazionale delle Ricerche (CNR), in Rome \cite{Colantoni}.

The \SI{90}{GHz} KID has a critical temperature $T_{c}=\left(812\pm 24\right)\SI{}{mK}$ and the noise equivalent temperature on the detector is ${\rm NET}=\left(6.94\pm 0.73\right)\SI{}{mK}/\sqrt{\rm Hz}$ \cite{Paiella}. The \SI{150}{GHz} KID has a critical temperature $T_{c}=\left(1.32\pm 0.04\right)\SI{}{K}$ and the noise equivalent temperature on the detector is ${\rm NET}=\left(0.909\pm 0.095\right)\SI{}{mK}/\sqrt{\rm Hz}$.

\subsubsection{Cryogenic system}
KIDs are low temperature detectors, they need to be cooled below the critical temperature of the superconducting film in order to work. The optimal choice is at least $T\lesssim T_{c}/6$.

The cryogenic system is a three-stage cryostat composed of a pulse tube cryocooler, a $^3$He/$^4$He fridge, and a dilution refrigerator. This system is able to reach a base temperature of \SI{136}{mK}, under an optical loading of about \SI{14}{\micro W}, for about \SI{7}{hours}.

%In order to minimize the radiative thermal load, at each plate at \SI{45}{K}, \SI{3.5}{K} and \SI{1}{K} an aluminum shield is anchored. 

%The optical system inside the cryostat is composed of a window, a system of two lenses, a chain of four filters, and the focal plane, anchored to the \SI{1}{K} plate with 6 low conductivity fiberglass rods and in thermal contact with the mixing chamber, through a golden copper braid. The window and the two lenses have been fabricated in HDPE. One of them is mounted on the \SI{45}{K} shield, while the other is mounted on the support of the focal plane.

\subsection{Measurements}
\begin{figure}[!h]
\begin{center}
{\includegraphics[scale=0.45]{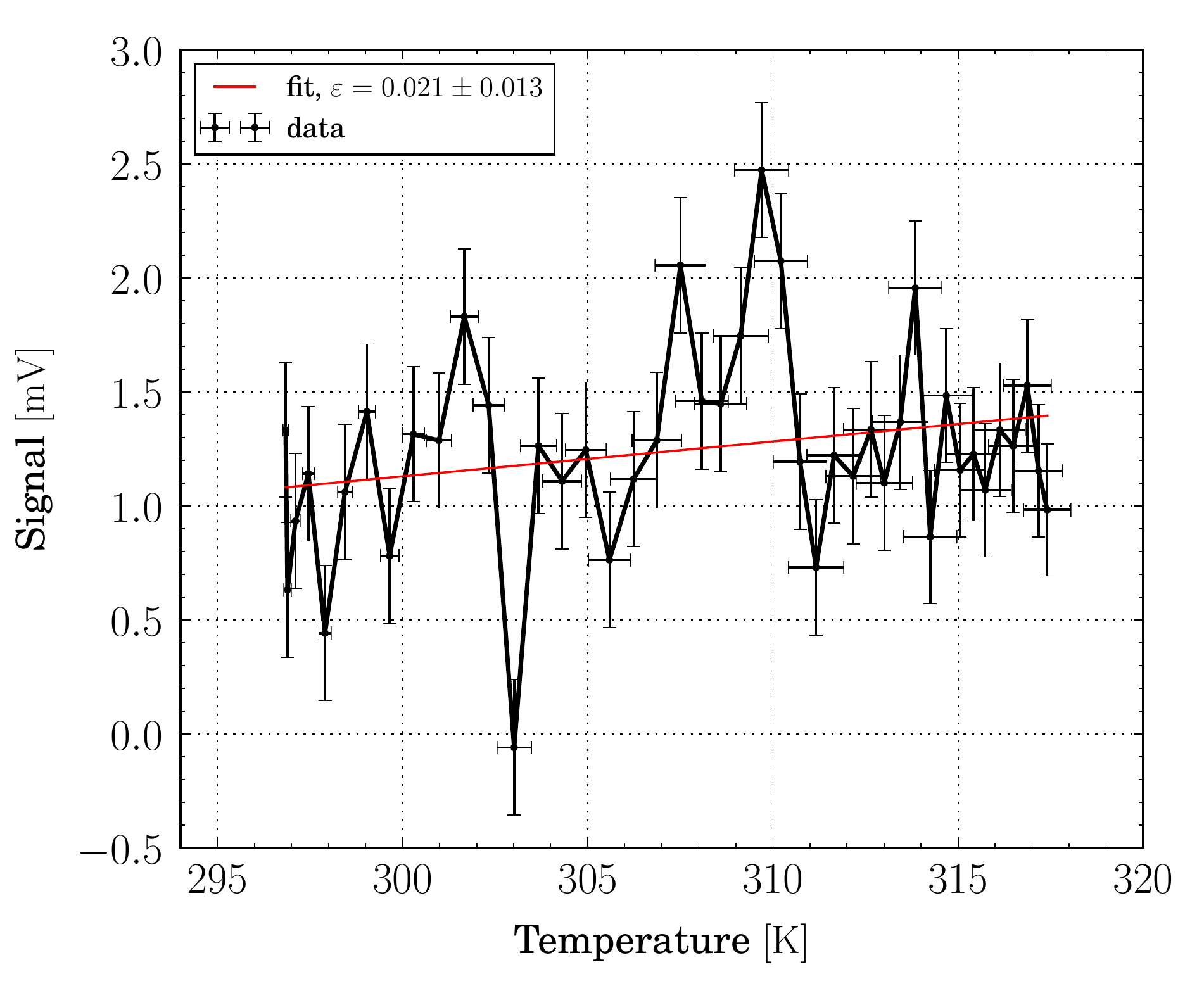}}\\
{\includegraphics[scale=0.45]{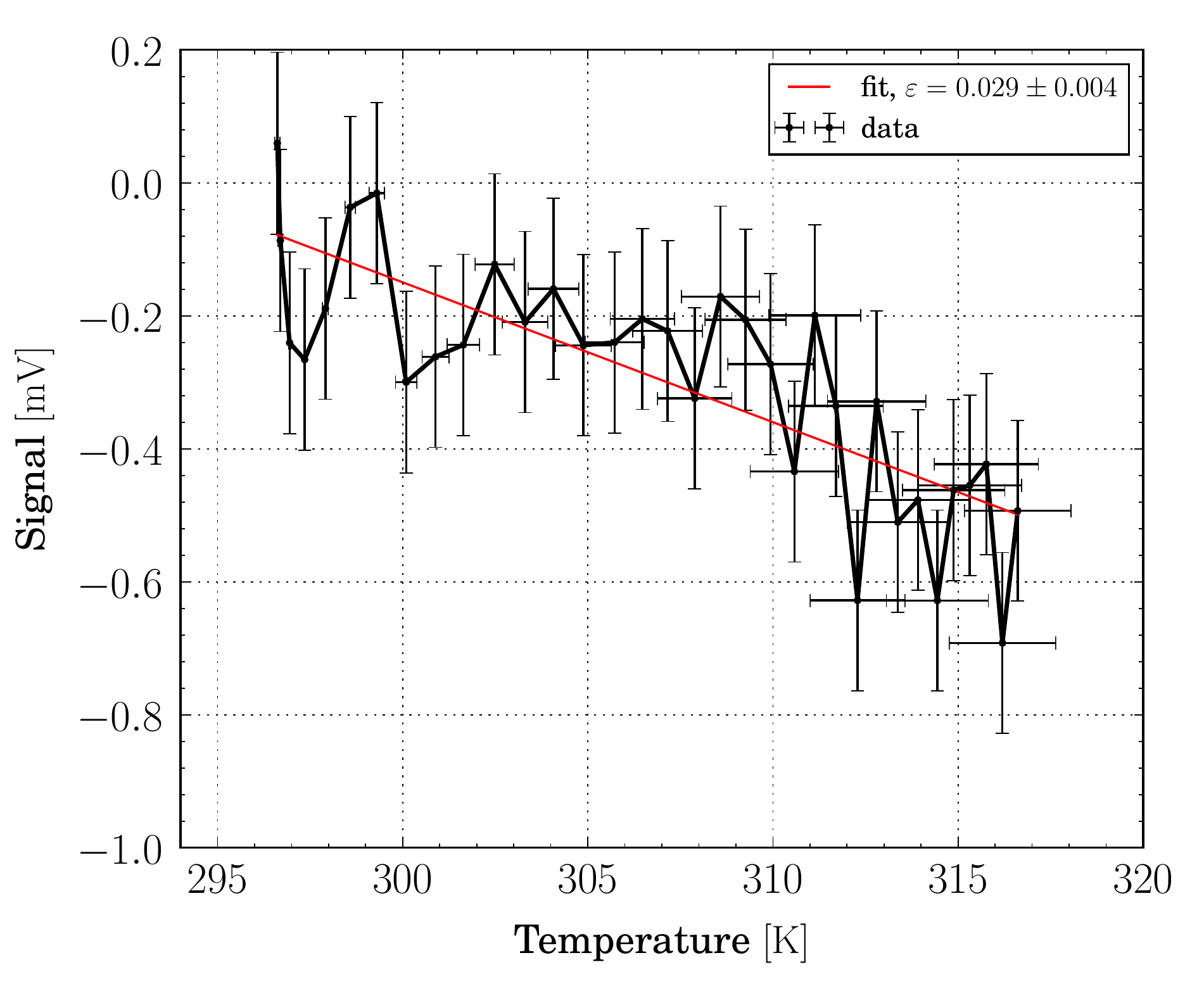}}
\caption{Measured intensity as a function of the UHMW temperature for the \SI{90}{GHz} (\emph{top panel}) and \SI{150}{GHz} (\emph{bottom panel}) detector.}
\phantomsection\label{misure}
\end{center}
\end{figure}

For the setup described before, the signal is 
\begin{equation}
S\!=\!\mathcal{R}\left\{\epsilon^{}_{\rm ecc}\left[T_{\rm amb}\left(t^{}_{\rm UHMW}+r^{}_{\rm UHMW}\right)-T_{\rm cho}\right]+\epsilon^{}_{\rm UHMW}T^{}_{\rm UHMW}\right\}
\phantomsection\label{eq:S}
\end{equation}
were $\mathcal{R}$ is the system responsivity in V/K, while $t^{}_{\rm UHMW}$ and $r^{}_{\rm UHMW}$ are the UHMW transmissivity and reflectivity, respectively, $\epsilon_{ecc}$ is the Eccosorb emissivity. The product $\mathcal{R}\epsilon^{}_{\rm ecc}$ is calibrated by removing the UHMW and placing an Eccosorb plate, cooled at $T_{N}=\SI{77.8}{K}$, behind the chopper, in this case the signal is
\begin{equation}
S_{\rm cal}=\mathcal{R}\left[\epsilon^{}_{\rm ecc}\left(T_{N}-T_{\rm cho}\right)\right]\;.
\end{equation}
The Eccosorb emissivity is estimated by measuring the transmissivity and the reflectivity of an Eccosorb slabs, using a high-power source: a \SI{150}{GHz} Gunn oscillator. We obtained $\epsilon^{}_{\rm ecc}=0.972 \pm 0.002$. At this point, the UHMW emissivity, $\epsilon^{}_{\rm UHMW}$, is measured by fitting the trend of $S/\mathcal{R}$ with the UHMW temperature, $T_{\rm UHMW}$, eq.~\eqref{eq:S}, see \figurename~\ref{misure}. Tab.~\ref{tab:fit} collects the values of the responsivities and the results of the fits found for the two detectors working at \SI{90}{GHz} and \SI{150}{GHz}. The integration time are around \SI{1}{hour} for both measurements.

\begin{table}[h]
\centering
\begin{tabular}{c|c|c|c|c}
\hline
\hline
f $\left[\SI{}{GHz}\right]$ & $\mathcal{R}$ $\left[\SI{}{mV/K}\right]$ & $\sigma_{\mathcal{R}}$ $\left[\SI{}{mV/K}\right]$ & $\epsilon$ $\left[\SI{}{\%}\right]$ & $\sigma_{\epsilon}$ $\left[\SI{}{\%}\right]$\\
\hline
\hline
90 & 0.72 & 0.01 & 2.1 & 1.3\\
150 &0.723 & 0.005 & 2.9 & 0.4\\
\hline
\hline
\end{tabular}
\caption{Values of the system responsivities and results of the linear fit on the trend of $S/\mathcal{R}$ with the UHMW temperature, for the two detectors working at \SI{90}{GHz} and \SI{150}{GHz}.}
\phantomsection\label{tab:fit}
\end{table}
%There is no physical reason to have two different emissivity at \SI{90}{GHz} and \SI{150}{GHz}, the measurements are indeed compatible at $1\sigma$, and we can define a mean emissivity as follow: 
%\begin{equation}
%\overline{\epsilon^{}_{UHMW}}=\frac{\left(\frac{1}{\sigma_{150}}\right)^2 \epsilon_{150} + \left(\frac{1}{\sigma_{90}}\right)^2 \epsilon_{90}}{\left(\frac{1}{\sigma_{150}}\right)^2+\left(\frac{1}{\sigma_{90}}\right)^2 }=2.89^{+0.5}_{-2.0}\%\;.
%\end{equation}
%The error is calculated as the maximum dispersion in order to be conservative. 

\section{Transmission}
The transmission at millimeter wavelengths is a crucial property because it discriminates if a material can or can not be used. A proper material, for both the cryostat window and the lenses, must have a transmission near \SI{100}{\%}, in such a way that the optical losses, through these elements, are negligible. Nowadays, the most used material, which satisfies this requirement, is the HDPE \cite{Lamb, 1985ApOpt..24.4489H}. 

We performed transmission measurements on HDPE and UHMW samples with different thicknesses, and without any anti-reflection coating treatments.   

\subsection{Instrumental setup}
For this measurement, the instrumental setup is composed of a Martin Puplett Interferometer (MPI), a Hg-Lamp (blackbody at \SI{4000}{K}) and an Eccosorb sheet (blackbody at \SI{300}{K}) as sources, and a Golay cell as detector The Golay cell has a quartz lens which is a low-pass filter at \SI{1}{THz}. The Wire Grids used for the MPI starting to be ideal at \SI{90}{GHz} so we decided to show the measurement from \SI{150}{GHz} to \SI{870}{GHz}.

In the inputs of the MPI, we placed the Hg-Lamp and the Eccosorb sheet at room temperature, while the samples are placed directly in front of the Golay cell. The input signal is modulated at \SI{3.6}{Hz} through a chopper, and it is collimated on the detector through a Quartz lens, which acts as a \SI{1}{THz} low-pass filter as well. The output signal is demodulated through a Lock-in amplifier.

\begin{figure}[h]
\begin{center}
{\includegraphics[scale=0.25]{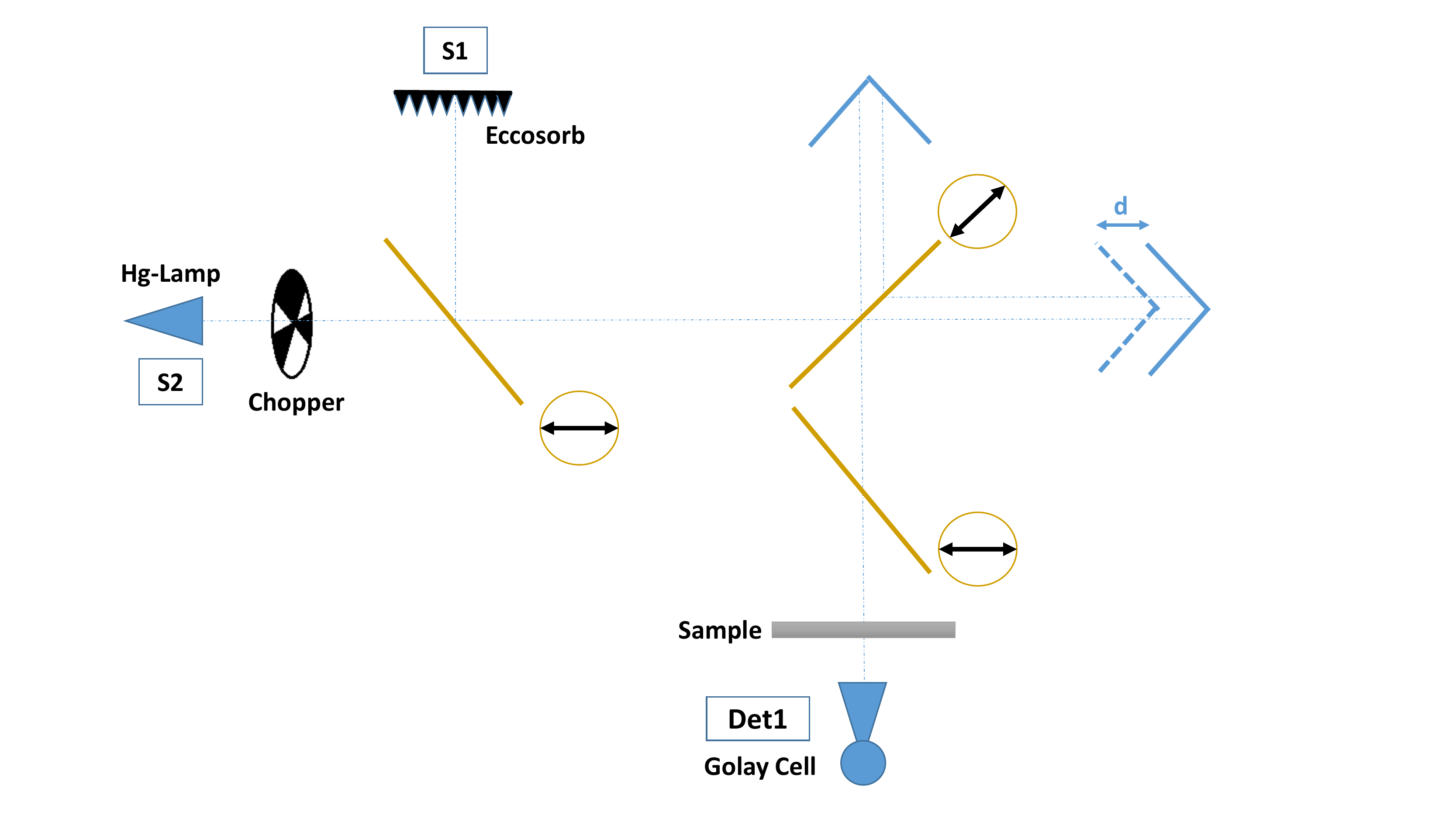}}
\caption{Scheme of the measurement setup used for the transmissivity and for the refraction index. The sample is placed between the last wire grid of the MPI and the detector, for the measurement of the transmission, while it is placed between the beam splitter and one of the two roof mirrors, in order to increase the phase shift, for the measurement of the refraction index.}\label{mpi}
\end{center}
\end{figure}

\figurename~\ref{mpi} shows a scheme of this setup, which is the same used in \cite{common_dale}, in such a way that all the systematics are under control \cite{2013InPhT..58...64S,common_dale,beam_splitter_error,daleSPIE}.
\subsection{Measurements}
In order to measure the transmissivity, we need to perform two measurements: with and without the sample. For each measurement, we acquired a single interferogram, around \SI{15}{minutes} of integration time, and we calculate the associated power spectrum by using a discrete fast Fourier transform algorithm. The spectrum obtained with the sample is normalized to that obtained without the sample. Assuming that the power produced by the emission of the sample is negligible, the intensity, $I_0$ , to the detector is 
\begin{equation}
I_0\left(\nu\right)\left[t^{}_{\rm UHMW}\left(\nu\right)+r^{}_{\rm UHMW}\left(\nu\right)\right] \ll \epsilon^{}_{\rm UHMW}\cdot BB\left(T=\SI{300}{K},\nu\right)\;,
\end{equation}
were $BB\left(T=\SI{300}{K}\right)$ is the black body intensity at room temperature, $\nu$ is the frequency, $t_{UHMW}$, $r_{UHMW}$ are the transmissivity and reflectivity of UHMW sample respectively, we can write: 
\begin{equation}
\begin{split}
\frac{S\left(\nu\right)}{S'\left(\nu\right)}&=\frac{I_0\left(\nu\right)\left[t^{}_{\rm UHMW}\left(\nu\right)+r^{}_{\rm UHMW}\left(\nu\right)\right]}{I_0'\left(\nu\right)}=\\
&=t^{}_{\rm UHMW}\left(\nu\right)+r^{}_{\rm UHMW}\left(\nu\right)\;.
\end{split}
\end{equation}

\begin{figure}[h]
\begin{center}
{\includegraphics[scale=0.5]{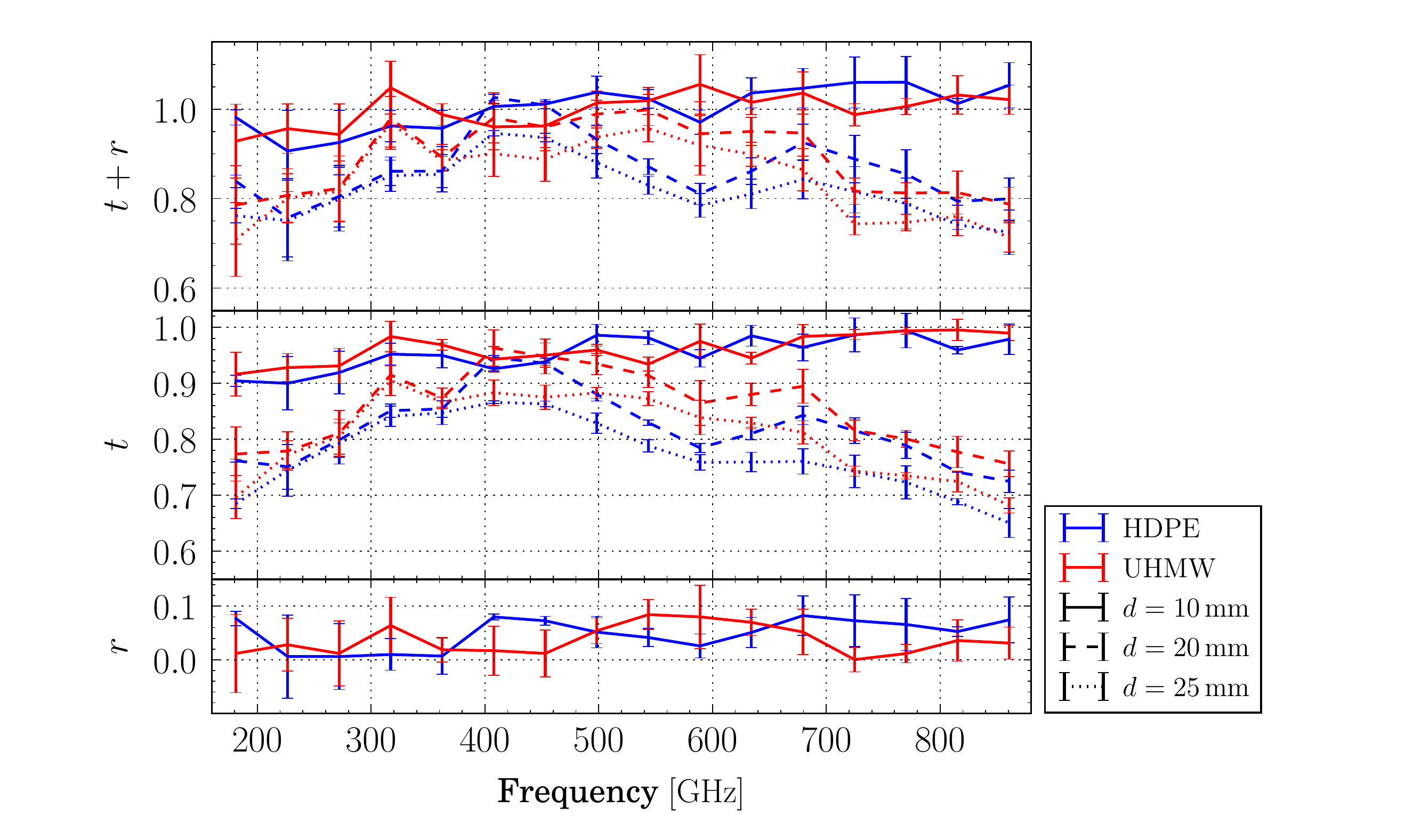}}
\caption{Transmissivity (\emph{central panel}) and reflectivity (\emph{bottom panel}) for the samples of HDPE (\emph{red}) and UHMW (\emph{blue}). The transmissivity is shown for different thicknesses of the samples: 10 (\emph{solid line}), 20 (\emph{dashed line}), and \SI{25}{mm} (\emph{dotted line}). The \emph{top panel} shows the trend of the sum of the transmissivity and the reflectivity.}
\phantomsection\label{f1}
\end{center}
\end{figure}

In order to discriminate the transmissivity from the reflectivity, we can use the thickness dependence of the transmissivity for each frequency:
\begin{equation}
\frac{S}{S'}=t^{}_{\rm UHMW}+r^{}_{\rm UHMW}=\left(1-{\rm e}^{-d_0/d}\right) +r^{}_{\rm UHMW}\;.
\phantomsection\label{eq:s/s'}
\end{equation}
For each frequency, we performed a two-parameter exponential fit of $S/S'$ as function of $d$, eq.~\eqref{eq:s/s'}, obtaining the values of $d_0$ and $r^{}_{\rm UHMW}$. We found a reflectivity less than the \SI{10}{\%} for both HDPE and UHMW, and a transmissivity greater than the \SI{90}{\%} for the \SI{10}{mm} thick samples of both materials. \figurename~\ref{f1} shows the transmissivity and the reflectivity for the different samples of HDPE and UHMW. From the top panel of \figurename~\ref{f1} it is possible verify $$ r(\nu)+t(\nu)+\sigma_r(\nu) + \sigma_t(\nu) < 1 \ \ \ \ \ \ \forall \nu  $$
so the energy conservation isn't violated.
 
%\subsection{Residual polarization}
Since the UHMW is formed by long chains of polyethylene, all aligned in the same direction, the presence of a  small residual polarization is possible. Performing different transmissivity measurements by rotating the sample, We did not obtain an evidence of this effect due. %to the sample and we can put an upper limit to this residual at few percent. 
However we are preparing dedicated measurements about this. A small effect due to polarization in polymeric materials has been already highlighted in \cite{Coppi2016}.

\section{Refractive Index}
The last optical property needed for the design of the lenses is the refractive index, which can be measured through a MPI by looking for the \emph{zero path difference} (ZPD) shift due to the passage in a different material. The optical shift $\Delta x$ is proportional to the thickness of the material, $d$, according to the following equation: 
\begin{equation}
\Delta x = \left(n-1\right) d\;,
\phantomsection\label{eqindex}
\end{equation}
where $n$ is the real part of the complex refractive index $\widehat{n}$:
\begin{equation}
 \widehat{n}  = n - ik,
\phantomsection\label{eqindex2}
\end{equation}
we can express the imaginary part as a function of absorption parameter $\alpha$:
\begin{equation}
 \alpha = 4\pi f k/ c,
\phantomsection\label{eqindex21}
\end{equation}
where $f$ is the frequency, $c$ is the light speed and $k$ is the extinction coefficient.

The HDPE refractive index at millimeter wavelengths is $1.54$ \cite{Lamb, 1985ApOpt..24.4489H,BIRCH1981225}. The absorption coefficient in the same range at 300K is \SI{0.03}{Np\cdot cm^{-1}}\cite{Lamb, 1985ApOpt..24.4489H,BIRCH1981225}. 

\subsection{Instrumental setup}
The setup is the same described for the transmission measurement, with the sample located between the beam splitter and one of the two roof mirrors. In this way, there is an additional phase shift $\Delta x$ and, therefore, a different ZPD position. We perform the measurement with \SI{10}{cm^{-1}} band-pass filter. 

\subsection{Measurements}
\begin{figure}[!h]
\begin{center}
{\includegraphics[scale=0.45]{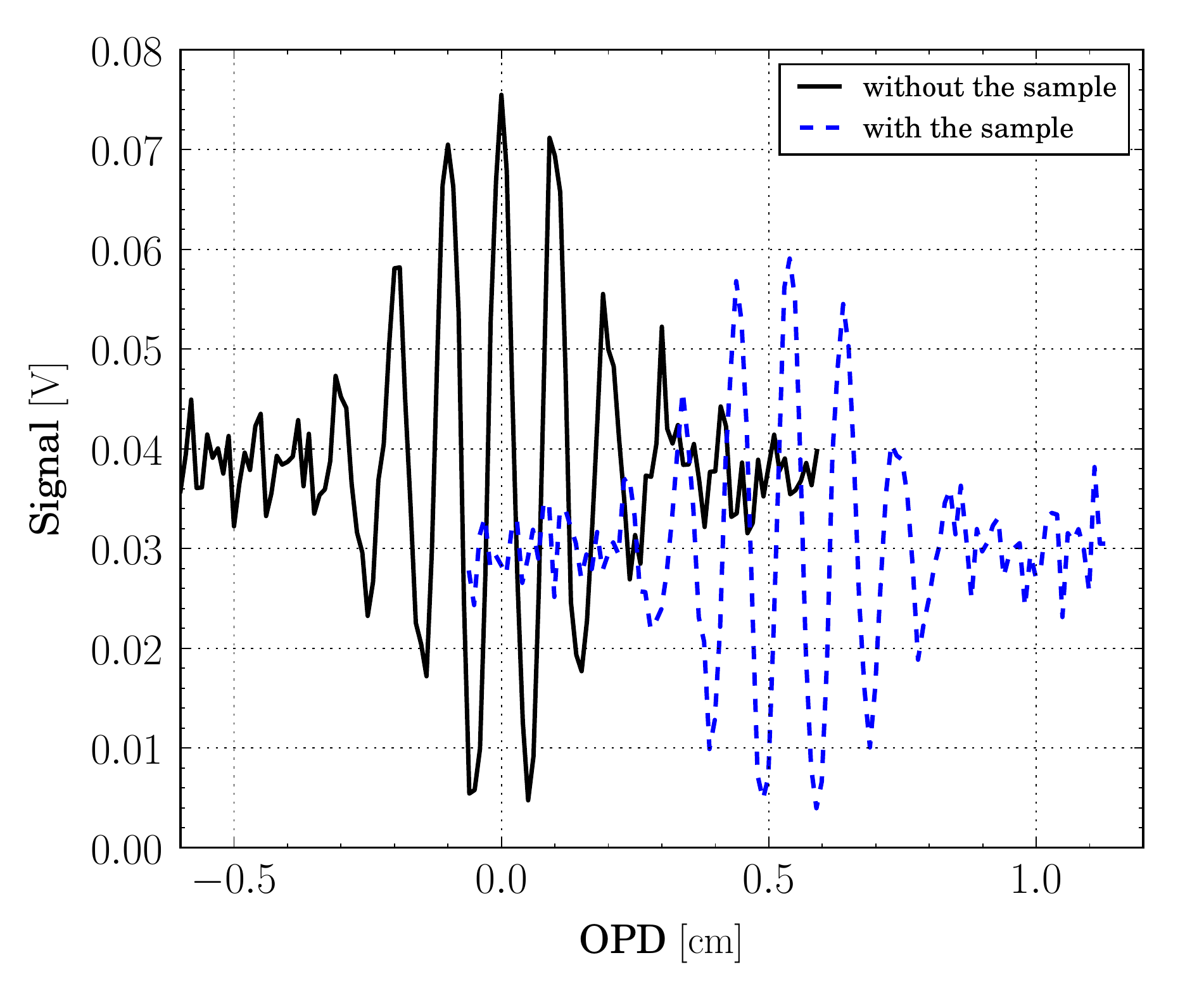}}
\caption{Raw interferogram produced by the difference of two blackbodies (\SI{4000}{K} and \SI{300}{K}) by using a MPI. The signal is plot as a function of Optical Path Difference (OPD). The \emph{solid black line} represents the interferogram measured without the sample, while the \emph{dashed blue line} represents the interferogram measured with the \SI{10}{mm} thick sample of UHMW placed on the delay line. The different position of the ZPD is directly dependent on the refractive index of the material according to eq.~\eqref{eqindex}.}
\phantomsection\label{interfero}
\end{center}
\end{figure}

We performed a number of measurements on three different thicknesses of the UHMW sample. In \figurename~\ref{interfero} is shown a raw measurement of the \SI{10}{mm} thick sample; each interferogram is performed by around 15 minutes of integration time. From preliminary measurements, we acquired long interferograms, finding raw ZPD positions, then we refined the measurements around the ZPD positions, in order to improve the precision. Tab.~\ref{tab:results_ZPD} summarizes the results for the ZPD shifts and the refractive indexes for the different thicknesses of the UHMW sample, we show also an upper limit for the absorption factor. 

\begin{table}[!h]
\centering
\begin{tabular}{c|c|c|c|c|c}
\hline
\hline
Thickness $\left[mm\right]$& $\overline{\Delta x}\left[cm\right]$& $\sigma_{\Delta x}$ & n& $\sigma_n $ & $\alpha[Np\cdot cm^{-1}]$ \\
\hline
\hline
10 & 0.539& 0.006& 1.539 & 0.008 &  $0.03$\\
20 & 1.075 & 0.03 & 1.54 & 0.02 &  $0.04$\\
25 & 1.33 & 0.03 & 1.53 & 0.02 &  $0.03$\\
\hline
\hline
\end{tabular}
\caption{Results for the ZPD shifts and the refractive indexes for the different thicknesses of the UHMW sample at \SI{300}{GHz}.}
\phantomsection\label{tab:results_ZPD}
\end{table}
Finally, we found the following mean refractive index at \SI{300}{GHz}: 
\begin{equation}
n= 1.537 \pm 0.009\;.
\end{equation}
and the upper limit for the absorption coefficient at \SI{300}{GHz}:
\begin{equation}
\alpha=  0.03 \pm 0.01 \ \ Np\cdot cm^{-1}\;.
\end{equation}
The results are compatible with the HDPE \cite{Lamb, 1985ApOpt..24.4489H,BIRCH1981225} and it is compatible with the results obtained in \figurename~\ref{f1} at \SI{300}{GHz}. 
\section{Conclusion}
We perform dedicated measurement in order to put in light the goodness about this material for millimeter wavelength applications. We measured its emissivity at \SI{90}{GHz} and \SI{150}{GHz} obtaining few percent, this results are consistent with similar plastic polymer material as the HDPE. By using a MPI and a Golay Cell, as detector, we measured the transmissivity as a function of the thickness. We measure $\sim 90\%$ from \SI{150}{GHz} to \SI{800}{GHz} for ten millimeter sheet. From these measure we derive the surface reflectivity, without coating, always less $20\%$. The results are very similar to HDPE. By using the same MPI as before we measured the refractive index, $1.537$, and  the absorption coefficient, $0.03Np \cdot cm^{-1}$, both obtained at \SI{300}{GHz}. 
The optical feature, very similar to HDPE, and the better mechanical characteristic make this material very suitable for millimeter wavelengths.  

\section{Founding}
This work has been supported by ASI (Agenzia Spaziale Italiana), grants OLIMPO and Millimetron, by PNRA (Italian National Antarctic Research Program), and by Sapienza University of Rome \emph{research-startup} funds.

\section{Acknowledgments}
We warmly acknowledge Mr. Giorgio Amico for careful machining of many part of the experiment.

\section*{References}
% Bibliography
\bibliography{biblio2}

\end{document}